%% file: main.tex
\documentclass{vgtc}                          % final (conference style)
\ifpdf%                                % if we use pdflatex
  \pdfoutput=1\relax                   % create PDFs from pdfLaTeX
  \usepackage{graphicx}                % allow us to embed graphics files
  \DeclareGraphicsExtensions{.pdf,.png,.jpg,.jpeg} % for pdflatex we expect .pdf, .png, or .jpg files
\else%                                 % else we use pure latex
  \usepackage{graphicx}                % allow us to embed graphics files
  \DeclareGraphicsExtensions{.eps}     % for pure latex we expect eps files
\fi%

\graphicspath{{figures/}{pictures/}{images/}{./}} % where to search for the images

\usepackage{microtype}                 % use micro-typography (slightly more compact, better to read)
\PassOptionsToPackage{warn}{textcomp}  % to address font issues with \textrightarrow
\usepackage{textcomp}                  % use better special symbols
\usepackage{mathptmx}                  % use matching math font
\usepackage{times}                     % we use Times as the main font
\usepackage{cite}                      % needed to automatically sort the references
\usepackage{tabu}                      % only used for the table example
\usepackage{booktabs}                  % only used for the table example
\onlineid{0}

\vgtccategory{Research}

\vgtcinsertpkg

\preprinttext{To appear in an IEEE VGTC sponsored conference.}

\title{Real-time Collaboration Between Mixed Reality Users in Geo-referenced Virtual Environment}

\author{Shubham Singh\and Zengou Ma\and Daniele Giunchi\and Anthony Steed}
\affiliation{\scriptsize University College London}

\abstract{

Collaboration using mixed reality technology is an active area of research, where significant research is done to virtually bridge physical distances. There exist a diverse set of platforms and devices that can be used for a mixed-reality collaboration, and is largely focused for indoor scenarios, where, a stable tracking can be assumed. We focus on supporting collaboration between VR and AR  users, where AR user is mobile outdoors, and VR user is immersed in true-sized digital twin. This cross-platform solution requires new user experiences for interaction, accurate modelling of the real-world, and working with noisy outdoor tracking sensor such as GPS. In this paper, we present our results and observations of real-time collaboration between cross-platform users, in the context of a geo-referenced virtual environment. We propose a solution for using GPS measurement in VSLAM to localize the AR user in an outdoor environment. The client applications enable VR and AR user to collaborate across the heterogeneous platforms seamlessly. The user can place or load dynamic contents tagged to a geolocation and share their experience with remote users in real-time.

} % end of abstract

\CCScatlist{
  \CCScatTwelve{Human-centered computing}{Human computer interaction (HCI)}{Interaction paradigms}{Mixed/augmented reality}{};
  
  \CCScatTwelve{Human-centered computing}{Visualization}{Visualization application domains}{Geographic visualization}{};
  
}

\begin{document}

%% The ``\maketitle'' command must be the first command after the
%% ``\begin{document}'' command. It prepares and prints the title block.

%% the only exception to this rule is the \firstsection command
\firstsection{Introduction}
\maketitle

\input{introduction}

%% \section{Introduction} %for journal use above \firstsection{..} instead

\section{Related Works}
\input{relatedworks.tex}

\section{System}
\input{system.tex}

\section{Testing}
\input{testing.tex}

\section{Evaluation}
\input{evaluation.tex}

\section{Limitations and Future Works}
\input{limitationsAndfutureworks.tex}

\section{Conclusion}
\input{conclusion.tex}

%% if specified like this the section will be committed in review mode
\acknowledgments{
We are thankful to the CS department of UCL for provisioning us with all the equipments and lab space for the development of this project.}

\bibliographystyle{abbrv-doi}

\bibliography{main}
\end{document}

%% file: introduction.tex
% Suggestion: Topic->context->background->problem & challenges->how your research tackles the problem->contribution

%The current state of the art mixed reality (MR) development allows one to present hyper-realistic 3D visualization of virtual objects and avatars. Using MR we can do spatial mapping for the understanding of physical space, estimating real-world lighting for illumination and shadows of virtual objects, gesture recognition for free interaction with virtual objects, etc. In virtual reality (VR), the user is completely immersed in a virtual world and isolated from the real world. In virtual reality, the user can experience depth of field - selective area of focus, foveated rendering using eye-tracking, 200 degrees of field of view (diagonally) with 110 Hz of refresh rate, tracking of rigid bodies with 6 degrees of freedom (DoF), etc gives hyper-realistic experience. The current state of the art VR systems are comparable to the human visual system. This is to attest to the fact the MR has developed to the point where users can expect to have an enhanced immersive and seamless virtual and mixed reality experience.

Collaboration is arguably the most prominent potential applications of mixed reality (MR) and virtual reality (VR) technologies. It allows one to connect with geographically (or physically) distant users immersively using the graphical interface of the computer. Some of the common use-cases of collaboration in mixed reality includes - virtual training, teleoperations, telexistence, immersive communication, etc. \cite{6,27,Tachi_telexistence_1985}. One important point to note is the collaborative mixed-reality is an asymmetric experience, the visualization and experience of virtual content varies between MR and VR user. The MR presents a combination of the virtual environment and the real world in three-dimensional space to the user, whereas, in VR, the user is completely immersed in the virtual world and blocked from the real-world. For a collaborative system, this poses a variety of unique challenges, like - freedom of movement, method of interactions, synchronization across platforms, etc.

%This makes it ideal for collaborative tasks, learning, and entertainment purposes. 
%We have seen several single-user applications of MR in the field of - medical- 3D visualization of human anatomy, industrial - assembling and disassembling of machinery, tourism - virtual 3D environments or 360 video experiences, etc.

%With the advancement of technology and the availability of affordable consumer devices, we have seen a proliferation of MR devices, and it's application in recent years.

The remote collaboration between users across the platform is an active area of research and development in virtual reality \cite{6,25,26,27}. Most of the recent collaboration efforts are centred for indoor experiences and homogeneous platforms (either mixed or virtual reality). For example, collaboration in "Spatial" \cite{spatial} is targeted for mixed-reality devices (i.e. MS-Hololens) only, and Microsoft "Azure spatial anchors" \cite{azure_anchor} supports multiple platforms, but strictly ties the content to the local scanned space of users. Restricting the scale of experience to local environment and inaccessible for VR users. 

The collaboration across heterogeneous platforms that support AR and VR is challenging, in the sense that the AR platform has an added dimension of the real-world along with virtual. The AR users are constrained by real-world conditions such as freedom of movement, spatial obstruction/distractions, size of the virtual model, etc. Whereas in VR, the entire world is simulated and independent of users' physical space. Users in VR can adapt to the virtual environment and its different factors such as the speed of movement, size of avatars, etc. \cite{28}. During the collaboration, the user in VR can not see the physical world of the AR user or their environmental state, like - traffic signals, events, etc. The remote VR user is oblivious to physical obstructions around the AR user, which might disrupt their experience. These differences may also pose accessibility issues for AR users, where the virtual content is not reachable. Other than accessibility and environmental differences, there are also challenges with visual cues and sense of agency. VR users can move and navigate much faster in the virtual world as compared to AR users in the physical world, this would pose a problem of tracking visual cues for AR users. Furthermore, the design guidelines for creating mixed reality experience is very diverse and lack standardization \cite{5}. 

%In this project, we aimed to minimize the differences in user experience of the platform and reduce the friction to interact and collaborate seamlessly across platforms. 

%Furthermore, the human-computer interaction in VR and AR is an active research area. Multiple ways of implementing the same feature exist. Some of the design decision that we considered were full-body avatar representation for remote user, real-world size map scale and 3D buildings for AR user, settings to switch between spatial and 2D audio, option for VR user to change moving speed for exploring the larger area, map alignment for AR users to adjust their position and rotation, 360 photospheres for visualizing real photos of the place, etc. 

Further, there have been efforts to implement AR or MR experience with spatial understanding \cite{27,spatial,34}, but, it usually focuses on the understanding and mapping of physical world space rather than localising the user on a geographical space. We have found only one study on the applications of collaborative MR in geographical context \cite{35}. Thus it seems that the geographical context and large scale application of Mixed reality is largely unexplored. The extended scale MR application can unlock new potential for the technology, which will allow the users to create and consume information/contents on geographical location/scale and also share it with others. The geolocation will add more gravity to information, and immersive 3D content will provide enriched ways of interactions.

Accurate location tracking of the user is necessary for real-time collaboration on such a large scale environment. The inaccurate tracking of sensors and misalignment of information contents will result in a very disconnected experience and undesirable glitches. For VR users, since everything is simulated virtually, we can accurately determine the users' position with respect to the virtual map and place virtual contents over the map accurately. But in the case of the AR user, we need to rely upon the Global Positioning System (GPS) tracking or Visual Inertial Simultaneous Localization and Mapping (VISLAM) system to localize the AR user. GPS alone is not reliable for accurate tracking of the users' motion, and VISLAM is not suitable for large scale applications due to scaling and accumulation of errors. To obtain accurate and reliable results, VISLAM requires detecting enough distinct features and accurate motion detection from onboard device sensors like accelerometers and gyroscopes. For a real-time collaboration application, the inaccuracy in the tracking of AR users may result in poor effects such as spatial jittering of users' avatar and false positional values, and other inconsistencies in synchronization of experiences. These issues will break the collaboration with other users in the session.

To implement collaboration between users on different platforms, we have included the following features - voice conferencing, freehand drawing in 3D space, and some dynamic experiences (like - 360 photosphere and guided tours). Although these features are common for MR applications and in collaboration, we present our approach to implement these features with geosynchronous context. For instance, the relevance of 360 Photosphere gets enhanced when placed at a geographical location. This gives a portal-like experience to the AR user while providing realistic geographical information for VR user. Further, we demonstrate a working solution to reinforce the GPS accuracy using ARCore VSLAM system for smooth tracking of AR user and synchronization with remote users for collaboration.

%To make the application independent of platform we have used Unitys' generic XR development, which allowed us to support HTC VIVE, Oculus and WMR headsets for VR client application. For AR client application we have used ARCore SDK, which we can use for Android and IOS platform. 

%% file: relatedworks.tex
%In this section, we are going to discuss on three topics. GPS and VISLAM are 

In this section, we will discuss four topics related to our work - GPS, VISLAM, hybrid methods and collaboration. We will first introduce GPS and VISLAM respectively and then compare their plausibility in the context of collaboration using MR. Then, we will explore studies that tried to combine different technologies to enhance the AR tracking system. Many user experience (UX) and human-computer interaction (HCI) studies are trying to explore which factors affect the experience and how to naturally present the virtual/synthetic information to both VR and AR users collaboratively. Our project is motivated by related research works. 

\subsection{GPS for AR tracking}
GPS is an earth-orbiting satellite-based navigation system \cite{9}. Due to the scale and the accessibility, GPS is the main choice as an outdoor tracking system for mobile devices. For example, Thomas et al. \cite{34} in 2002, demonstrated their outdoor AR gaming system - ARQuake, where they have implemented tracking for the AR system using Differential GPS (DGPS).

However, using GPS tracking will result in an offset and continuous fluctuation by several meters away from the true location. Benford et al. \cite{35} showed that in their game, which used GPS and Wifi for tracking users' positions, the GPS resulted in errors ranging from 1m to 384m. There are multiple reasons for GPS inaccuracy, and they are inevitable. Selective availability is a primary source error, which was purposely imposed by the US Government to restrict the accuracy of the GPS \cite{10}. Some errors are introduced by the atmosphere, like the Ionospheric delay and Tropospheric delay. Some ascribe to the surrounding around the GPS receivers, like the multipath interference, the superposition of multiple GPS signals, or even the attenuation by the concrete \cite{9, 13}. Another complicating factor is that the satellites are in non-stationary orbit, the user standing at the same location will get varying GPS signals \cite{gps_variability_steed_2004}. These error sources result in poor GPS accuracy.

Researchers are attempting to make the GPS more reliable and accurate. Mobile phones are typically using Assisted GPS (A-GPS) to increase sensitivity and signal availability. Information, like GPS health and ionospheric model, can be used to help detect GPS signals by calculating the frequency of each satellite due to Doppler shifting \cite{11}.

Various signal receivers or processors may store or transmit GPS values with different decimal places. It can cause rounding errors from a few centimetres to a couple of meters. Because of the precision errors, the virtual objects will be placed with offsets so that they can be beyond the physical boundaries. Thus, they may not be visible or accessible to AR users. Also, the alignment or offset problems will confuse GPS based synchronization and collaboration between VR and AR users. Therefore, the GPS signals alone are not sufficient for AR, especially for cross-platform collaboration. 

\subsection{VISLAM for AR tracking}

Visual Inertial SLAM (VISLAM) augments the SLAM (or GPS) using external sources like inertial measurement unit (IMU) to localize the sensor/camera. If manually created landmarks of the environment or GPS signals are available, the advance computer vision methods may not be required to improve the accuracy. For example, ARQuake used GPS for most of the time, but if there were landmarks on the wall, the system would switch to use the landmarks as they provided more accurate locations \cite{34}. However it's not practical to place landmarks (fiduciary markers) manually in different physical environments, thus we need more automated and robust method. VISLAM provides an alternative with the absence of preloaded maps in the mobile devices \cite{14}. VISLAM addresses the method of creating a map of an environment from a continuous sequence of images  \cite{16, 36}. Depth information is calculated by using different images of the same scene with distinct feature points \cite{17, 18}. Also, information from the Inertial Measurement Unit (IMU) sensors, like accelerometers and gyroscopes \cite{19}, is used to determine the camera positions and orientations in real-time and improve the noisy estimations with Kalman filter \cite{3}. VISLAM is a robust tracking \cite{2, 22} for indoor environments with stable light conditions and sufficient feature points. Therefore, for most of the AR application, VISLAM is an ideal approach. Apple and Google released their AR toolkit (ARKit and ARCore respectively) for developers using this method for tracking.

VISLAM alone is not sufficient to localize the user/camera in geospatial coordinate. Since positional tracking is relative to the starting point of the local user and does not take geolocation into account, we can not use it for cross-platform collaboration that requires a global coordinate system. Thus we look for hybrid solutions to combine the GPS scale with VISLAM precision.

\subsection{Hybrid of GPS and VISLAM}
Over the last decade, the progress in VISLAM has enabled outdoor tracking \cite{1, 2, 19, 20, 21}. Nonetheless, it is still challenging to use VISLAM as an outdoor tracker for AR applications as it suffers from issues like - dynamic environment, changing illumination, scaling errors, etc. Researchers are trying to leverage the strength of GPS and VISLAM to get a better result for outdoor AR tracking.

Vineet and Amir \cite{24} demonstrated the use of GPS signals and 3 degrees of freedom (DOF) orientation sensors to track AR users’ outdoor positions and demonstrated CAD model in an outdoor environment. Schleicher et al. \cite{29} showed the improvement of outdoor GPS tracking by combining with SLAM system. These research work has inspired us to combine GPS and SLAM, which provides 6 DOF (both translation and orientation). Daniel and Todd \cite{16} showed the plausibility of this idea by fusing bundle adjustment-based VSLAM and differential GPS, to obtain precise GPS and altitude measurement. In this paper, we prototyped a similar idea by using VSLAM of ARCore and device GPS to get a cheaper and more accessible solution.

\subsection{Cross-Platform Collaboration}
In present days, digital communications and long-distant teleconferencing have become the norm for personal and professional communications. However, there is still a huge gap between audio-video (AV) conferencing and face-to-face communications, since people not only use speech but also different gestures, gaze and other non-verbal cues to express their ideas and thoughts \cite{6}. To show more explicitly the expressions, poses or movement, avatars can be used in virtual telepresence applications. In research studies, participants have reported a much stronger sense of presence in virtual conferencing than traditional audio or video conferencing \cite{7, 25}. In the last decade, the shared virtual environment has developed significantly and seen larger adoption from people. Kiyokawa et al. \cite{26} initiated a novel idea which allowed cross-platform collaboration between AR and VR. Our design will expand in this area by creating an application that can bring remote people into a common geographic location for a collaborative experience.

% Piumsomboon et al. \cite{27} pointed out several methods to enhance AR and VR interactions with gestures, head rotation, and eye gaze. They stated that virtual cues could improve awareness of remote collaborators, and the size of the avatar could influence the collaborative behaviors.  Interestingly, they pointed out that snapping VR users’ avatar to AR perspective could enhance the collaboration. Their paper supported our ideas. We synchronized events for both VR and AR users. Users with different platforms can watch videos, see the virtual geo-tags of the building, draw and see others’ drawings in real-time. For VR users, head and eye gaze can be added to further enrich the immersive experience. 

In previous studies, many collaborations were carried in a way that one user, who could see an overall view, helped the other with an egocentric viewpoint to do some tasks \cite{30, 31, 32}. From the previous works, people can identify the factors that will affect users' performance in cross-platform collaboration. Piumsomboon et al. \cite{27} pointed out several methods to enhance AR and VR interactions with gestures, head rotation, and eye gaze. They stated that virtual cues could improve awareness of remote collaborators, and the size of the avatar could influence the collaborative behaviours. Similarly, Oda et al. \cite{32} used the advantage of the virtual cues to convey instruction of assembly more effectively. 

Our application provides egocentric viewpoints to the users on each platform. Instead of one user giving a set of instructions to the others, we allowed users to experience the same scenario without the restriction of their locations or any physical difficulties. In another study of adaptive avatars Piumsomboon et al. \cite{33} pointed out that, with redirected gaze and gestures, the performance of remote collaboration can be enhanced as non-verbal information can be easily conveyed. Hence, complex animation of the avatars, head and eye gaze of the VR users can also be added to enrich the immersive experience further and support collaboration. 
%We synchronized events for both VR and AR users. Users with different platforms can watch videos, see the virtual geo-tags of the building, draw and see others’ drawings in real-time. This collaboration should reinforce the natural social cues as we tried to minimize the distinction of the experience between them.
%VR and AR platforms. 

\subsection{Digital Twins}

With the accelerated progress in the field artificial intelligence (AI), internet of things (IoT), virtual reality (AR or VR) and big data, there's no dearth of visionary projects like - virtual Oulu \cite{virtual_oulu_2016}, Paris 2.0, digital build Britain \cite{digital_britain}, etc. Where the efforts are underway to combine these technologies to create a digital model of any city, also referred to as "Digital Twin" \cite{digital_twin_2019}. Madni et. al. \cite{digi_twin_def_madni_2019} define the digital twin as - \begin{quote}
A Digital Twin is a virtual instance of a physical system (twin) that is continually updated with the latter’s performance, maintenance, and health status data throughout the physical system’s life cycle.
\end{quote}

The virtual reality (VR/AR) is a perfect medium for visualization and consumption of such huge amount of data intuitively. The Digital twin model is currently being used for providing an extensible 3D interface of the city, virtual collaboration, geographical information system (GIS), etc. \cite{virtual_oulu_2016}

%Based on the previous studies, more features can be added in a later version. In our case, we disabled the avatar for the local user and showed it only for remote users now. This design choice was made for the consistency of VR and AR experience. However, Steed et al. \cite{7} in 2016 pointed out that self-avatar would enhance a user’s cognition in the virtual environment and provide more direct information in manipulation and communication. 

%% file: system.tex
Our goal is to build an application that allows AR and VR users to collaborate synchronously in real-time in the same geographical location. In this section, we will describe the implementation and architecture of our solution and motivate the design decisions that we have made during the development.

\subsection{Design}

To get the geographical data, we needed a map service that can be used on multiple platforms (Windows and Android for this project) and flexibility to control the map data and visualization. Mapbox provides an easy to use SDK with features, like – customizable tilemaps, layers for map data, procedural building and textures, feature modifiers, etc. which make it ideal for our solution. For multiplayer networking we have using Photons' real-time networking solution called "photon unity networking" (PUN). Photon provides easy to use SDK for Unity3D and free to use cloud server for up to 20 concurrent users.

%While exploring different map SDK, such as – Google Maps, Scape, and Mapbox, we found Mapbox to be most suitable for our requirement. Mapbox has easy to use SDK with relevant features, like – customizable tilemaps, layers for map data, procedural building and textures, feature modifiers, etc. The Mapbox service is hosted remotely on the public cloud, and the map data and layers can be accessed using a public API access token. All the map data and layers can be independently managed from a web dashboard and Mapbox studio, which is accessible from our user account.

%For cross-platform collaboration, we decided to go with Photons’ real-time networking solution called “Photon Unity Networking (PUN)”. We have used PUNs’ self-hosted license and deployed the server on windows server 2019 Datacenter on Azure. Our server can manage up to 100 concurrent users with a latency of about 30 ms (max) for UK-region. The server facilitates a multiplayer framework with cross-platform support for Windows, Android, iOS, and Web platforms. We have used this server for synchronous collaboration between users across platforms. The same server is also used for voice conferencing between users using a VoIP. All the current communication between users is broadcasted over an unencrypted binary protocol with a reliable-UDP connection.

\begin{figure}[!tb]
 \centering % avoid the use of \begin{center}...\end{center} and use \centering instead (more compact)
 \includegraphics[width=\columnwidth]{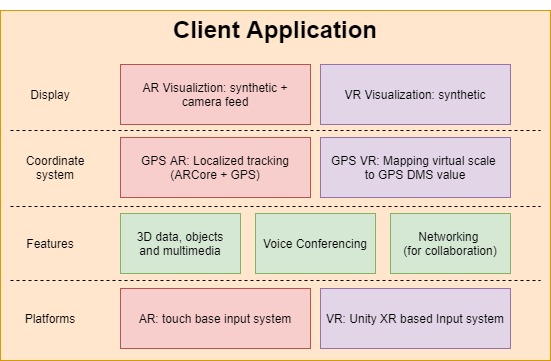}
 \caption{Client application architecture (Combined representation for AR and VR). Blocks are color coded, where red is for AR platform, Blue for VR platform and green is common across platforms.}
 \label{fig:architecture}
\end{figure}

The architecture of AR and VR client applications is shown in \autoref{fig:architecture}. The key differentiation between the two client applications is in terms of the tracking of geolocation and visualization. The AR user sees the virtual map environment and information overlaid on top of the live camera feed. In contrast, the VR user is completely immersed in the virtual world and cut-off from the physical world.
%The tracking of users in AR and VR is completely different. 
For AR users, we use the corrected GPS position with respect to the local map coordinate system
%, which is then continuously synchronized over the network with remotes users
For VR users, we can directly determine the users’ position in local virtual map coordinates. Apart from these key architectural differences, there are other lower level differentiation for input systems and interaction for different features of the application.

\begin{figure}[!tb]
 \centering
 \includegraphics[width=4cm, height=9cm]{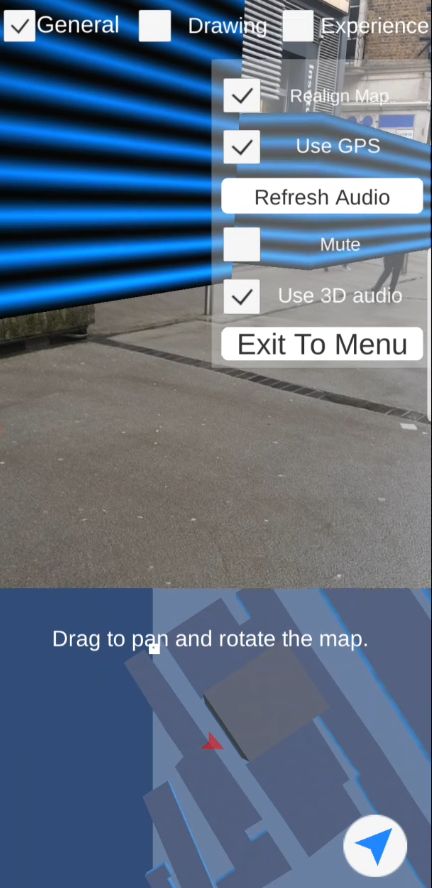}
 \includegraphics[width=4cm, height=9cm]{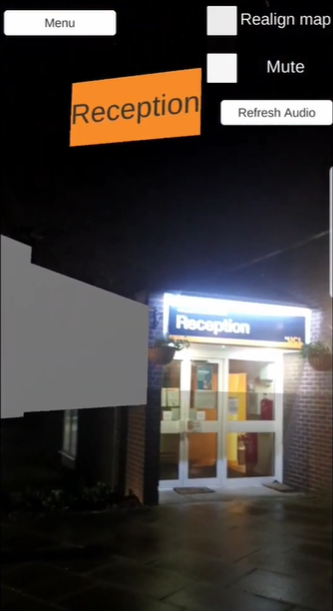}
 \caption{Menu settings for virtual map alignment (left) and geo-tagged label (right) in AR}
 \label{fig:map_align}
\end{figure}

We have used Mapbox service to create virtual tilemaps and 3D procedural world. Mapbox procedurally generates 3D buildings, textures, and vegetation using depth and layer information from world map (taken from open-source data such as - OpenStreetMap and NASA). It then places these 3D features over tilemap with respect to geolocation and assigns a unique ID for each building. It uses primitive polygons (like cube and triangles) to construct the buildings and environment, which may not be desirable for a context-aware application. The procedurally generated objects and buildings can be replaced with custom models and prefabs using the "feature modifier" option from the SDK. We have replaced the primitive buildings with our own realistic 3D models to create a more realistic experience and provide geographical landmark cues.

% Voice settings
The default voice setting for the user is set to 3D (Spatial) so that they can have a spatial understanding of remote users and have a more immersive experience during collaboration and group tasks. Further, the audio can be switched to 2D mode, which allows users to communicate irrespective of their virtual distances with everyone else.

\subsection{Static GPS correction}
Map alignment is an essential step for correcting the registration of AR users for global positioning (GPS). Since the GPS reading can vary from 1m to 384m \cite{35} depending upon location and coverage, its raw value is not usable for real-time collaboration. Thus manual alignment is essential step at the beginning of every session to correct the map positioning and orientation manually. The user can align the virtual map manually by using touch and drag input on the orthographic top-view, as shown in the \autoref{fig:map_align}. Once this is done, the tracking can be reliably done in an outdoor area. The alignment can be done in middle of session as well, in case the tracking gets wrong. 

%Our implementation of improving GPS precision relies upon the user to manually align the map direction and position at the beginning. Because there's no guarantee that the initial GPS value recorded from on-board system is correct, thus we need the user to manually correct the starting point so that it's close to the truth value. The user can align virtual map manually by using touch and drag input on the orthographic top-view, as shown in the \autoref{fig:map_align}. The default voice setting for user is set to 3D (Spatial), so that they can have spatial understanding of remote users and have a more immersive experience during collaboration and group tasks. Further, the audio can be switched to 2D mode, which allows users to communicate irrespective of their virtual distances with everyone else.

%Next, once the alignment is set, user can update the settings to start the SLAM reinforcement of GPS readings.For reinforcing GPS value from SLAM we linearly combine the last set GPS value (manually aligned by user) with regularly updated AR camera root transform (which is updated by ARCore) to smooth any fluctuations in GPS reading and then use the updated transform to estimate the new GPS value in local map coordinate system and synchronize this new value with remote users.

%% Implementation sub-section
\subsection{Dynamic GPS correction}
In previous sections, we have discussed on the manual correction of the GPS tracking for AR user and map alignment. The precise location tracking is very important for real-time interaction with remote users. To improve the GPS tracking, we have linearly combined the GPS value from the device with the VISLAM system of ARCore. GPS value from the device is used to determine the initial location of the AR user. Once the starting position of the AR user is set, the default GPS mode is disabled from settings to use the corrected-GPS mode. The ARCore regularly provides the virtual camera transform using visual SLAM. We take this updated transform and find its relative position from the maps starting point and calculate the corrected GPS positions for the virtual camera with respect to local map coordinates. This new corrected GPS value is then broadcast to all the participants in the session. For VR users, we can get the effective  GPS value from their virtual map coordinate system. The local user will determine their GPS value (corrected-GPS for AR user and virtual GPS for VR user) and broadcast it to all the participants in the session. The received GPS values from remote users are then converted into the local coordinate system for positioning and visualization. That's how users over different platforms and location synchronize the position of virtual objects and avatars using the GPS value.

%The precise location tracking is very important for cross-platform interactions. For AR user, we have combined the GPS value from device with the VISLAM system of ARCore. GPS value from device is used to determine the initial location of AR user. User can also manually adjust and align the virtual map with their physical environment for precise tracking. Once the starting position of AR user is set, the outputs from ARCore will update the corrected GPS values. For VR user, we can get the precise GPS value from their local coordinate system. Position and experience of both the AR and VR user is synchronized using the GPS value. The local user will determine their position value and broadcast to all the participants in session. The received GPS values are then converted into local coordinate system for visualization. In this way, we achieved the real-time collaboration between cross platform users in geographical context.

\begin{figure}[!tb]
 \centering % avoid the use of \begin{center}...\end{center} and use \centering instead (more compact)
 \includegraphics[width=4cm, height=9cm]{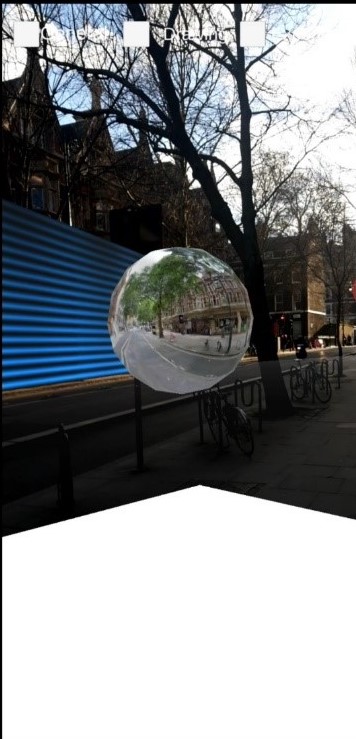}
 \includegraphics[width=4cm, height=9cm]{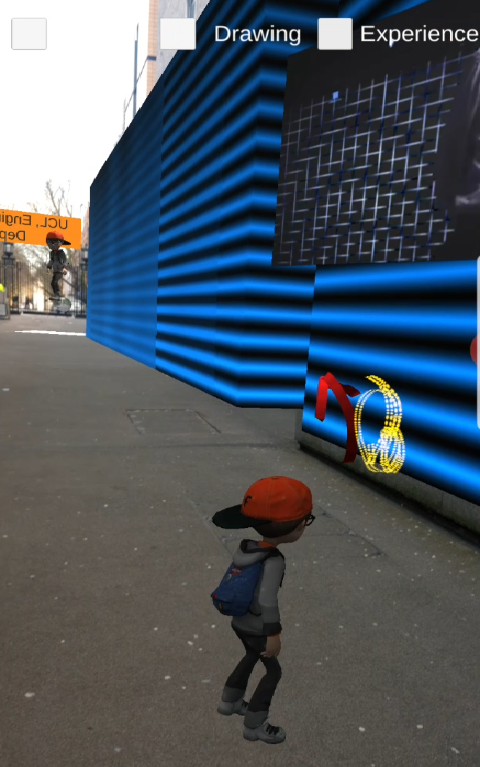}
 \caption{Interactive 360 Photosphere (left) placed at (51.522771, -0.131833). It holds a captured 360 image of the place. On right, we see multi-user collaboration, where 3 users are in session and watching a video displayed on right wall and drawing in 3D space.}
 \label{fig:photosphere}
\end{figure}

\subsection{User Avatar}
For cueing the presence of a remote user, we have implemented a full body avatar with animation states. Presently there are only two animation states of – Idle and walking, but this can easily be extended to include more intricate inverse kinematic (IK) and blendtree animations. The animation states are synchronized with remote users using remote procedure calls (RPCs). Steed et al. in 2016 pointed out that self-avatar would enhance a user’s cognition in the virtual environment and provide more direct information in manipulation and communication \cite{7}. Although in our case, we disabled the avatar body for local user and enabled it only for remote users. We decided to avoid implementing complex animation states for VR users and to keep it consistent with the experience of AR users, in which case we only have a virtual camera to control the avatar.

\subsection{Free hand drawing}
Normally, in collaboration applications, we observe that drawing/annotation is synchronized by the stroke or line basis. We have implemented the spatial drawing (in 3D space) using pointwise synchronization to give real-time visualization experience.  Drawing settings for both VR and AR users provides different options for brushes, size, and colour. Currently, we have used four different types of brush which varies in animation and shape.

For freehand drawing, we had two methods for implementation. First using the line renderer and second using a mesh-based approach. We followed the later one to get more control over the geometry of the line and using vertex shader for rendering different effects. For drawing in VR, the user press and holds the right trigger button and then moves the controller in 3D space freely \autoref{fig:ucl_quad}. In AR, the user touches and holds on the screen and moves the phone to draw in 3D space. For line drawing, we process the continuous stream of 3D position from VR/AR input system, which is used to create 3D mesh. Because of the platform differences and geographical environment between the users, we can not simply synchronize the point position. Doing so will result in distorted reconstruction due to GPS precision error, which is (~70 cm for 5 decimal precision of GPS values). To avoid this precision error, we only use the GPS value for the first point of the line stroke and for every next point we pass the relative vectors with respect to the first point of the stroke. So for a continuous input of 3D position, this relative position is synchronized with remote users to reconstruct the line locally in real-time. The dynamic texture on brush pre-set was implement using CG shader to optimize the performance impact.

%The line drawn is made from a continuous sequence of points, which are received from input sources (hand controller in VR and touch input in AR). Using these sequential points, we generate a connected quad mesh to create a solid surface, and the orientation of this quad depends upon the relative position of the current point from the last one. To synchronize the drawing, we only convert the first position of the point to the GPS values, which acts as a spawning position for the object. The rest of the points are synchronized relative to the starting point, and only the distance vectors are passed to avoid the GPS precision errors.

\subsection{Geographical cues}

For enriching the immersive experience, we have added an interactive 360-degree Photosphere. This Photosphere shows the real world captured image of the place; this allows the user to have a portal-like experience where they can step inside different experience that holds the 360-degree image of the place (\autoref{fig:photosphere}). The Photosphere reacts to the users' position, and scales up and down in proportion to the distance from user. So if the user is approaching the Photosphere it scales up to cover their view and reaches maximum scaling once the user is at its center. These 360-degree images were downloaded using “StreetViewDownload 360” software \cite{street_view_sw}, which is a freeware to download the 360 images from Google street view using URL ID for offline use.

\begin{figure}[!tb]
 \centering
 \includegraphics[width=9cm]{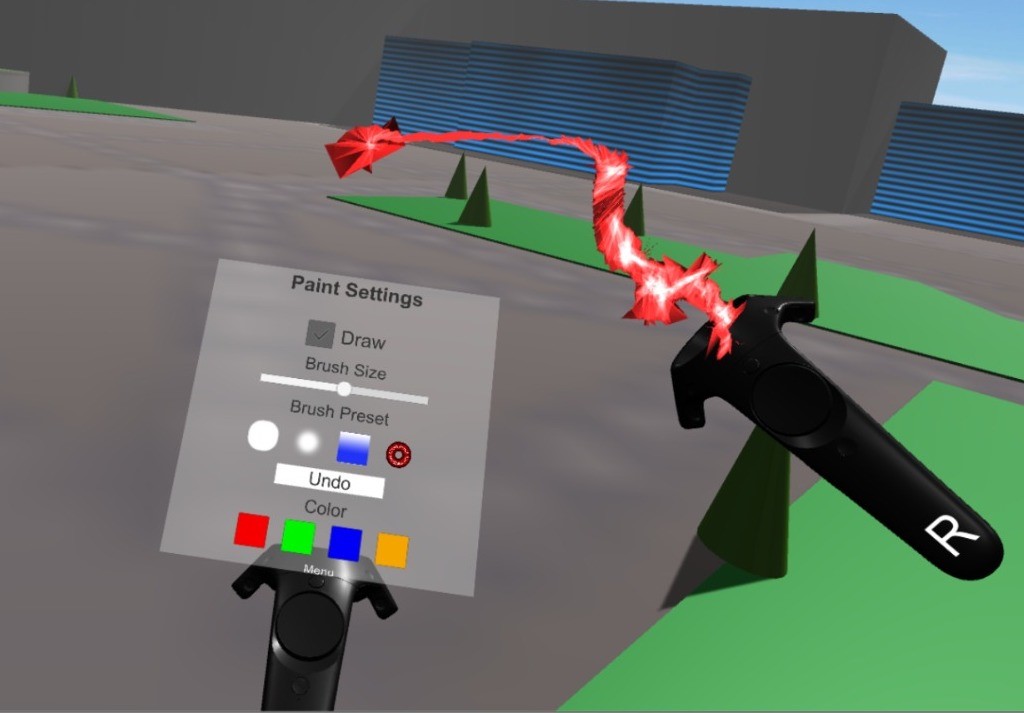}
 \includegraphics[width=9cm]{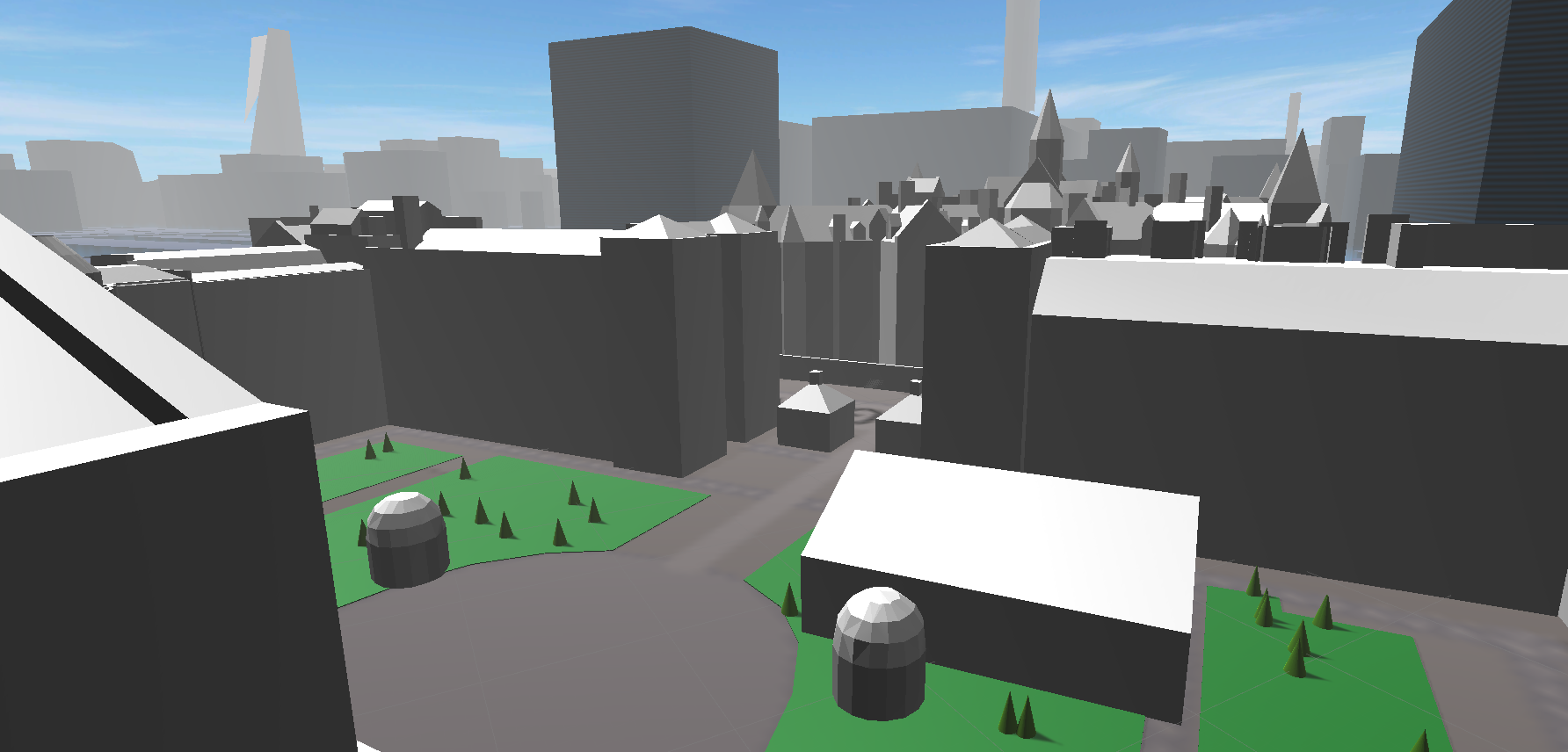}
 \caption{Spatial drawing and settings (Top) and 3D environment of UCL main quad (Bottom) in VR.}
 \label{fig:ucl_quad}
\end{figure}

To enrich the virtual experience further we have used realistic 3D models of UCL campus and nearby area. The procedural buildings of Mapbox include primitive polygons and do not include any structural details or information. Fortunately, they do have scope for feature modification in SDK, which makes it easier to replace procedural buildings with custom 3D models. To replace any procedural building, we need to block the primitive buildings by their unique ID from spawning and replace it with our model with appropriate map scale. Presently our source UCL model does not include any texture. So we have used diffuse white materials for building. But this can easily be improved by involving 3D artists or buying premium models online. See \autoref{fig:ucl_quad} for reference virtual environment.

%% file: testing.tex
%To support our development, we have thoroughly tested both the client applications (AR and VR). The users were able to create and join new session. Multiple users from different platforms can join the same session and collaborate, we have tested with a maximum of 3 users in a session (2 in VR platform and 1 in AR platform). For testing our client applications, we have used a VR-ready Windows desktop machine and Android phone (API level 28) with ARCore support. After joining the session, the users can use voice conferencing, avatar, and drawing tool to collaborate with remote users. We tested this by drawing collaboratively in same geolocation space and controlling dynamic experiences for remote users, where the host of the session can load new experiences. 

%We did a quantitative analysis of the location tracking system. To test the improvement of our corrected GPS tracking for AR users, we planned the area using Google maps and identified suitable closed paths. The objective was to test the loop closing for corrected SLAM system and GPS values, and compare the results. During the test, we simultaneously record the GPS reading and the corrected GPS values from our application on an Android device. A comparison between these two data sets was made by plotting the graph on the same diagram (see \autoref{fig:James} and \autoref{fig:Ifor}). The similarity between the two measurements and smoothness of the paths is observed to determine the better results.

We did a quantitative analysis of the location tracking system. For testing the correction of GPS reading using ARCore SLAM, we selected three outdoor paths in two different areas. We planned the area using Google maps and identified suitable closed paths. Two locations were around “St James’s Square, London SW1Y 4LE” (see \autoref{fig:James}). The third location was around “Ifor Evans Hall, Camden Road, London NW1 9HZ” (see \autoref{fig:Ifor}). The objective was to test the loop closing for corrected SLAM system and GPS values, and compare the results. During the test, we simultaneously recorded the GPS reading and the corrected GPS values from our application on the Android device. A comparison between these two data sets was made by plotting the graph on the same diagram (see \autoref{fig:James} and \autoref{fig:Ifor}). The similarity between the two measurements and smoothness of the paths is observed to determine the better results. The first two paths were used to test the performance with different areas and circumferences.

%For testing the correction of GPS reading using ARCore SLAM, we selected three outdoor paths in two different areas. Two locations were around “St James’s Square, London SW1Y 4LE” (see \autoref{fig:James}). The third location was around “Ifor Evans Hall, 109 Camden Rd, London NW1 9HZ” (see \autoref{fig:Ifor}). The first two paths were used to test the performance with different areas and circumferences. The map with the red line is the ground truth for the path that we have plotted on Google map. The other graph was plot from recorded GPS values, where on the x-axis, we plotted the longitude and on Y-axis the latitude. We understand that this 2D plotting is not accurate for spherical Geo-coordinate but we have assumed the distance difference to be negligible for such a small area.

%In testing, we found the rounding-off error of GPS values from Mapbox SDK. For instance, a label positioned at (51.522791, -0.131754) was shifted to (51.52279, -0.13175) in local map coordinates. The offset was calculated to be 29 cm away from the defined position, which is within 70 cm distance error, as we mentioned previously. Normally this error is acceptable in general GPS applications, but this error becomes relevant during collaborative drawing or for some contents located near the buildings. This error makes it hard for fine lines to be connected between different strokes of remote users. Although since we are doing point wise relative synchronization for stroke, the shape of each stroke remains intact, only the starting point gets effected.

%% file: evaluation.tex
%For testing the correction of GPS reading using ARCore SLAM, we selected three outdoor paths in two different areas. Two locations were around “St James’s Square, London SW1Y 4LE” (see \autoref{fig:James}). The third location was around “Ifor Evans Hall, 109 Camden Rd, London NW1 9HZ” (see \autoref{fig:Ifor}). The first two paths were used to test the performance with different areas and circumferences. The map with the red line is the ground truth for the path that we have plotted on Google map. The other graph was plot from recorded GPS values, where on the x-axis, we plotted the longitude and on Y-axis the latitude. We understand that this 2D plotting is not accurate for spherical Geo-coordinate but we have assumed the distance difference to be negligible for such a small area. 

The recorded values of onboard GPS and corrected-GPS were then compared to evaluate the efficiency of the method. For comparison, we plot these values on different map and graphs. The map with the red line is the ground truth for the path that we have plotted on Google map. The other graphs were plot from recorded GPS values, where on the x-axis, we plotted the longitude and on Y-axis the latitude. Then 500 points on each path are randomly selected for displaying (x) on the graph. The blue line represents GPS values recorded from the device and the orange line represents corrected GPS. We understand that this 2D plotting is not accurate for spherical Geo-coordinate but we have assumed the distance difference to be negligible for such a small area. 

\begin{figure}[!tb]
 \centering % avoid the use of \begin{center}...\end{center} and use \centering instead (more compact)
 \includegraphics[scale=0.7]{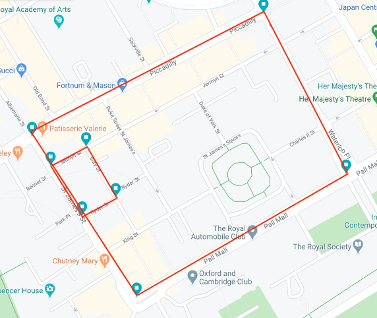}
 \includegraphics[width=4cm,height=4cm]{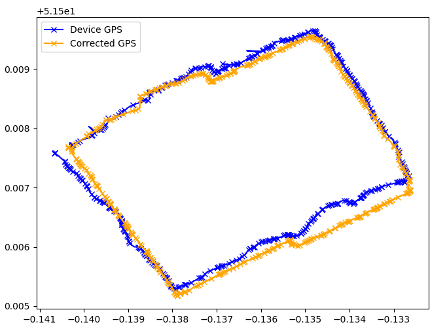}
 \includegraphics[width=4cm,height=4cm]{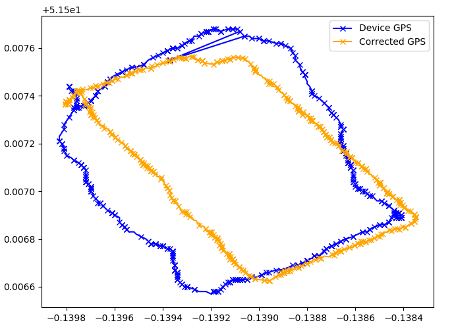}
 \caption{Path (large and small) around St James’ Sq. (Top) is shown on Google Maps and subsequently recorded GPS values for the large rectangular path on the left and smaller path on the right (Bottom) is plotted below. The blue line for device GPS and the orange line for corrected GPS.}
 \label{fig:James}
\end{figure}

\begin{figure}[!tb]
 \centering % avoid the use of \begin{center}...\end{center} and use \centering instead (more compact)
 \includegraphics[scale=0.7]{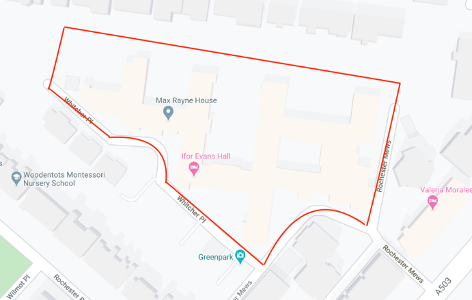}
 \includegraphics[scale=0.7]{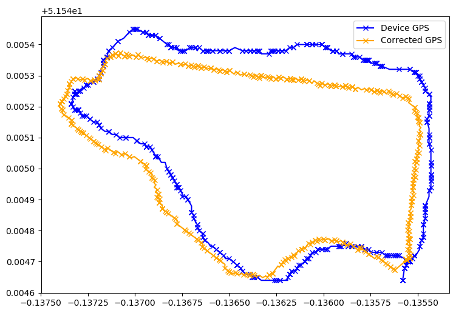}
 \caption{Path around Ifor Evans, UCL is shown on Google map and subsequent recorded GPS reading is plotted below. The blue line for device GPS and the orange line for corrected GPS.}
 \label{fig:Ifor}
\end{figure}

By inspection, the corrected GPS values were more smooth and less fluctuating, as compared to raw GPS values from the device. From the diagrams, the line segments of corrected GPS readings were straight and continuous, unlike the line segments of the raw GPS values. It could also be seen that the corrected GPS was closer to the ground truth value for all three paths. Thus, our approach provides more stable results compared to device GPS value. This corrected GPS values would be better for synchronization with remote users and collaboration. Consequently, we can also conclude that ARCore can be used to improve the GPS values for tracking AR users in small outdoor environments effectively (~136057 m sq. area of St. James test location). 

% added this from introduction - Shubham
Additionally, we measured error due to rounding-off of GPS coordinate from Mapbox SDK. It rounds off GPS value to 5 decimal places, whereas generally, 6 or more decimal places are used for accurate geolocation (i.e. in Google Maps). This rounding off of decimal places leads to a maximum round-off error of about 70 cm. Because 0.000005 degrees in latitude or longitude corresponds to 50 cm in distance, and it will result in about 70 cm diagonally away from the expected position. This may cause problems during the collaboration. For instance, a dynamic label positioned having geolocation of (51.522791, -0.131754) was shifted to (51.52279, -0.13175). The offset is calculated to be 29 cm away from the defined position, which is within 70 cm distance error, as we mentioned previously. 

%% file: limitationsAndfutureworks.tex
 
In this section, we will talk about the limitations of using ARCore for improving the accuracy of onboard GPS, some of the assumptions that we have made while developing this solution, and scope of this project to be extended for future research or applications.

In our implementation of tracking AR user, we have used ARCore but it's specifically targeted for indoor use cases and fails to continue tracking in larger outdoor areas. Arcore is not suitable for handling dynamic outdoor environments where you would expect moving crowds and objects. Among all existing popular AR systems, ARCore performed the worst result in a large and dynamic environments \cite{3}. Further development and customization in VISLAM technology may positively affect the improvisation of GPS reading in mobile phones. As an alternative to our implementation, we recommend working on customized and effective VISLAM, which can be used to maintain the ephemeral map of the virtual camera and can be used for reinforcing the GPS reading. Also, in general cases, the manual map alignment can be replaced by an automatic procedure and can be estimated by averaging initial GPS measures or relative sensor localization using VISLAM.

%The GPS coordinate rounding error will causes problems during the collaboration. If one user wants to do a specific operation at a certain location, the result may be show at a slightly different location in remote users' view. As we experience in case of freehand drawing. To compensate for these small offsets, we have used different threshold distances and relative positions to avoid flicker or unnatural behavior. Like in the case of switching between animation states or interacting with 360 photospheres.

The rounding off of GPS coordinates will causes problems during real-time collaboration. If one user wants to do a specific operation at a particular location, the result may be shown at a slightly different location in remote users' view. As we experience in case of freehand drawing. To compensate for these small offsets, we have used different threshold distances and relative positions to avoid flicker or unnatural behaviour. Like in the case of switching between animation states or interaction with 360 Photospheres. But alternatively, depending upon Map service this factor can be addressed differently.

Further, in the current implementation, we assume that the user is always at the ground level, and there's no account for their altitude or geographical elevation. Therefore, a possible problematic situation may happen if the user is trying to load an experience from indoor in a high building or using the application for navigation in hilly areas. Possible solution for this could be to use the terrain map, instead of flat tilemap to account for geographical elevation, and position the virtual experience accordingly.

We further plan to investigate and research on the following aspects - realistic 3D models for contextual information and immersion, psychological experiments in collaboration, and use cases of geographical data in mixed reality. Also, this projects' concept can be adapted for multiple industrial scenarios and use cases such as - tourism, urban planning, site inspection, training, treasure hunt, games based on geolocation, and much more. 

%% file: conclusion.tex
We have presented our implementation of mixed reality-based real-time collaboration application that supports multiple platforms. The application allows multiple users to share a common geographic location and have an immersive experience. We extended on different aspects of collaboration, to immerse and engage the users with virtual content by loading dynamic experiences, 360 Photospheres, and freehand drawing. This project serves as a valuable prototype for cross-platform collaboration between users in a geographical environment.

We have improved the GPS tracking for AR users using ARCore VISLAM system, which made the whole real-time collaboration possible. Stable and consistent tracking of remote users increases the sense of the presence for both AR and VR users and enhances the immersion of the user in an alternate reality/experience. The improved results are a good metric to further research in this area and explore more on collaboration use cases over geographical space. 

%In our tracking system, the drawback of using ARCore to modify GPS values is that ARCore is specifically targeted for indoor use cases, and it fails to continue tracking in larger areas and is unstable. It's not suitable for tracking in dynamic outdoor environment where you would expect moving crowd and objects. Among today’s popular AR systems, ARCore performed the worst result in a large area and dynamic environments \cite{3}. To expand the idea from this report, we recommend working on customized and effective VISLAM tailored towards the outdoor environment. 

%Further, the map service can be used and tailored for multiple industrial scenarios and use cases such as tourism, planning, site inspection, training, treasure hunt, games based on geo-location and much more. This project can also be extended for aspects like realistic 3D models for immersion, psychological experiments for collaboration, use cases of geographical data in mixed reality, etc. 